\documentstyle[twocolumn,eqsecnum,aps]{revtex}
\def\d{\dagger}
\def\a{\alpha}
\def\b{\beta}

\def\be{\begin{equation}}
\def\eq{\end{equation}}

\begin{document}
\draft
\title{The Influence of Higher Fock States in Light-Cone Gauge Theories }
\author{S. Dalley}
\address{Theory Division, CERN, CH-1211 Geneva 23, Switzerland}
\maketitle
\begin{abstract}
In the light-cone Fock state expansion of gauge theories, the influence
of non-valence states may be significant in precision 
non-perturbative calculations.
In two-dimensional gauge theories, it is shown how these states modify the
behaviour of the light-cone wavefunction in significant ways relative to
endemic choices of variational ansatz. Similar effects in four-dimensional
gauge theories are briefly discussed.
\end{abstract}
 
\narrowtext

\section{Introduction}

Light-cone quantisation is believed to be an especially efficient
hamiltonian method for dealing with many-body relativistic field theory 
\cite{rev}.
One of the tenets of this method is that the valence approximation
to a system of light-cone partons provides a much better approximation 
than the analogous Tamm-Dancoff truncation  in other quantisation
schemes. This chiefly follows from the peculiar inverse relation between
light-cone momentum and free energy:
\be
k^- = {m^2 + {\bf k}^2 \over 2 k^+} \ . \label{disp}
\eq
Here, $k^{\pm} = (k^0 \pm k^3)/\sqrt{2}$ are the light-cone momentum and
energy, ${\bf k} = (k^1 , k^2)$ the transverse momenta, and $m$ the mass
of a parton. Since $k^+ > 0$ and is conserved, to create more partons
costs significant energy because the new particles carry smaller $k^+$ than
their parent(s). Moreover, one might naively expect wavefunctions to vanish at
small $k^+$ to ensure finite energy.
In this report it will be shown how non-valence states nevertheless modify the
behaviour of light-cone wavefunctions, {\em including the valence sector},
when one or more parton momenta are small. Rather than wavefunctions 
simply  vanishing at the boundaries of phase space, the requirements of 
finite energy in general lead to cancelations between the valence and 
non-valence (sea)
components of boundstates. This phenomena was recently analysed for
QCD in ref.\cite{stan2}, where it was shown to be responsible, at least
perturbatively, for
the rising Regge behaviour of hadronic structure functions at small
momentum fraction $x$.
In this work some implications for non-perturbative calculations
will be demonstrated. 
The relevant effects may prove significant
in precision calculations with variational bases.
The basic physical ideas can be
illustrated by two-dimensional gauge theories. 
Implications for 
four-dimensional gauge theories will be discussed later.

\section{Two-Dimensional Gauge Theories}
The results to be discussed in this section will in fact be 
true for any two-dimensional gauge theory exhibiting
pair production. Rather than  
attempting to cover all particular cases at once, we proceed by using 
familiar examples. The generalisation will easily be seen to follow
from the simple structure  of two-dimensional gauge theories.
Therefore, consider the standard two-dimensional 
gauge theory with fermion fields $\Psi$  and action
\be
S= \int dx^0 dx^1  \ {\rm i} \overline{\Psi} \gamma_{\a} D^{\a} \Psi
 - m \overline{\Psi} \Psi - {1 \over 4 e^2} F_{\a\b}F^{\a\b}  \ ,
\eq
where representation and colour indices are suppressed for the moment, for
generality. Using a chiral representation of the two-dimensional $\gamma$
matrices in the
light-cone gauge $A_{-} = (A_0 - A_{1})/\sqrt{2} = 0$, we are left with
$A_+= (A_{0} + A_{1})/\sqrt{2}$ and the left-moving $\psi_{L}$ and
right-moving $\psi_{R}$ components
of $\Psi$. $A_+$ and $\psi_{L}$ are found to be constrained variables
with respect to light-cone time $x^+ = (x^0 + x^1)/\sqrt{2}$, and
can be eliminated by their equations of motion to yield
a light-cone hamiltonian 
\be
P^- = \int dx^- \ \  {m^2 \over 2} \psi_{R}^{*} {1 \over {\rm i} \partial_-}
\psi_R  + {e^2 \over 2} J^+ {1 \over ({\rm i} \partial_-)^2} J^+ 
 \ . \label{ham}
\eq
Here $x^{-} = (x^0 - x^1)/\sqrt{2}$ and $J^+$ is the right-moving
current $J^+ =  \psi_R \psi^{*}_{R}$. (The same form of hamiltonian results for
bosons $\phi$, with an appropriate current $J^+ = {\rm i} \phi^* 
\stackrel{\leftrightarrow}{\partial}_{-}
\phi$.) These theories generically exhibit linear confinement,
since the Coulomb potential is linear in one space dimension.
The hamiltonian (\ref{ham}) can be diagonalised in the light-cone Fock space
at fixed $x^+$, constructed from the Fourier modes of $\psi_R$,
\be
\psi_R (x^-) = {1 \over \sqrt{2 \pi} }
\int_{0}^{\infty} dk \ \left( b(k) {\rm e}^{-{\rm i} k x^-} +
d^{\d}(k) {\rm e}^{{\rm i} k x^-}\right) \ .
\eq
A (non-baryonic) 
eigenfunction of $P^-$ with total momentum $P^{+}$ 
is of the form 
\begin{eqnarray}
|\Psi(P^+)\rangle &  = & \int_{0}^{P^+} dk_{1} dk_{2} \ 
\delta ( P^+ - k_1 - k_2) \nonumber \\
&\times &\ f_2(k_1, k_2) b^{\d}(k_1) d^{\d}(k_2) | 0\rangle \nonumber \\
& & + {1 \over 2} \int_{0}^{P^+} dk_{1} dk_{2}  dk_{3}  dk_{4} \
 \delta \left( P^+ - \sum_{i=1}^{4} k_i \right) \nonumber \\
&\times & f_4(k_1, k_2, k_3, k_4) b^{\d}(k_1) b^{\d}(k_2) 
d^{\d}(k_3) d^{\d}(k_4) |0\rangle \nonumber \\
&& + \cdots \ .
\end{eqnarray}
Ellipses indicate higher Fock states $f_n$, $n > 4$.

Projecting $2P^+ P^- |\Psi(P^+)\rangle = M^2 |\Psi(P^+)\rangle$ onto 
specific Fock states, where the eigenvalue $M^2$ is the invariant mass
squared of the system, one derives coupled integral equations for the
wavefunction components $f_n$. Let us consider the 
familiar example of two-dimensional
QED (massive Schwinger Model at $\theta=0$ \footnote{Some observations
related to those made in this paper have recently been given in connection
with the $\theta$-dependence of the massive Schwinger model \cite{burk}}), 
which in the approximation that retains
only $f_2$ and $f_4$ reads \cite{berg,perry1}
\begin{eqnarray}
M^2 f_2(x,1- x)&  = & {m^2 \over x(1-x)}f_2(x,1- x)  \label{mass} \\ 
&& \! \! \! \! \! \! \! \! \! \! \! \! \! \! 
+{e^2 \over \pi} \int_{0}^{1} dy {f_2(x,1-x) 
- f_2(y,1-y) \over (x-y)^2} \label{cou} \\
&& \! \! \! \! \! \! \! \!  \! \! \! \! \! \! \! + 
{e^2 \over \pi} \int_{0}^{1} dy  f_2(y,1-y) \label{ano}\\
&& \! \! \! \! \! \! \! \! \! \! \! \! \! \! \! 
 + {e^2 \over \pi} \int_{0}^{1} dy_1 dy_2 dy_3 
\delta (y_1 + y_2 + y_3 - x) \nonumber \\
&& \times {f_4(y_1, y_2, y_3, 1-x) \over (x-y_1)^2}
\label{pair1} \\
&&\! \! \! \! \! \! \!  \! \! \! \! \! \! \! \!  
- {e^2 \over \pi} \int_{0}^{1} dy_2 dy_3 dy_4 
\delta (y_2 + y_3 + y_4 - (1-x)) \nonumber \\
&&\times {f_4(x,y_2, y_3, y_4) \over (1-x-y_4)^2} \ ,
\label{pair2} 
\end{eqnarray}
\begin{eqnarray}
M^2 f_4(x_1,x_2,x_3,x_4) & = & m^2 \sum_{i=1}^{4} {1 \over x_i}
f_4(x_1,x_2,x_3,x_4) \nonumber \\
&&  + {e^2 \over 2 \pi} A(x_1,x_2,x_3) \nonumber \\
&& ! \! \! \! \! \! \! \!  \! \! \! \! \! \! \! \! \! \! \! \!  
+ {e^2 \over \pi} \int_{0}^{1} dy_1 dy_2 
\left\{ \delta (y_1 + y_2 - (x_1 + x_2)) \right. \nonumber \\
&& \! \! \! \! \! \! \!  \! \! \! \! \! \! \! \!  \times
{f_4(y_1,y_2,x_3,x_4) - f_4(x_1,x_2,x_3,x_4) \over (x_1-y_1)^2}
\nonumber \\ && \left. + \ {\rm Similar} \  \right\} \ ,\label{four}
\end{eqnarray}
where and $x_i = k_i / P^+$, $x_4= 1- x_1 -x_2 -x_3$ and 
\begin{eqnarray}
A(x_1, x_2, x_3) & = & 
{f_2(1-x_3, x_3) -f_2(x_2,1-x_2)  
\over (1-x_2 - x_3)^2}  \nonumber \\
&& \! \! \! \! \! \! \!  \! \! \! \! \! \! \! \!  \! \! \! \! \! \! \!  
\! \! \! \! \! \! \!  \! \! \! \! \! \! \! \!  \! \! \! \! \! \! \!  
+ \ { f_2(x_1 + x_2+x_3, 1-(x_1 + x_2+x_3)) - f_2(x_1,1-x_1)
\over (x_2 + x_3)^2} \nonumber \\
&& \! \! \! \! \! \! \!  \! \! \! \! \! \! \! \!  \! \! \! \! \! \! \!
\! \! \! \! \! \! \!  \! \! \! \! \! \! \! \!  \! \! \! \! \! \! \!    
 + \ {f_2(x_2,1-x_2) - f_2(x_1 + x_2+x_3, 1-(x_1 + x_2+x_3))
\over (x_1 + x_3)^2} \nonumber \\
&&\! \! \! \! \! \! \!  \! \! \! \! \! \! \! \!  \! \! \! \! \! \! \!
\! \! \! \! \! \! \!  \! \! \! \! \! \! \! \!  \! \! \! \! \! \! \!    
+ \ {f_2(x_1,1-x_1) - f_2(1-x_3, x_3)
\over (1-x_1-x_3)^2}  \ .
\end{eqnarray}
The many `similar' contributions to the last term in (\ref{four}) are
enumerated in ref.\cite{perry1}, but are unimportant for our purposes.
At this point it is important to stress the distinction between the 
component $f_2$ to the solution of such equations in the valence
approximation, which sets $f_n = 0$ for $n >2$, and the component
$f_2$ to the solution when some or all of the higher Fock states
$f_n$ are retained. As is shown below, these have different high-energy
behaviour in general.

Like any wavefunction, the components $f_n$ are subject to 
high energy boundary conditions. From (\ref{disp}), this is a small
light-cone momentum boundary.
A typical variational ansatz used for the solution of the light-cone
boundstate integral equations
is
\be
f_n(x_1, \ldots, x_n) \sim 
x_{1}^{\b} x_{2}^{\b} \cdots x_{n}^{\b} \cdot P(x_1, \cdots, x_n) \label{ans}
\eq
where $x_n = (1- \sum_{i=1}^{n-1}x_{i})$ and  $P$ is polynomial. The 
safest procedure is to leave $\b>0$ as a variational parameter. However, often
the following formula is used
\be
m^2 = e^2 (1- \pi \b \cot{\pi \b}). \label{trans}
\eq
This choice is motivated by a study of the high-energy
`endpoint' behaviour of 
$f_n(x_1, \dots , x_n)$
as one of the arguments vanishes, $x_i \to 0$.
Ignoring terms resulting from a change
in the number of partons (pair production (\ref{pair1})(\ref{pair2})), 
one deduces that for finite
$M^2$,  as $x \to 0$ the divergent behaviour of the mass (\ref{mass})
and Coulomb terms (\ref{cou}) must cancel;
\begin{eqnarray}
&\lim_{x \to 0}
\left\{ {m^2 \over x}f_2(x,1- x)  + {e^2 \over \pi} 
\int_{0}^{1} dy {f_2(x,1-x) 
- f_2(y,1-y) \over (x-y)^2} \right\} &  \nonumber \\ 
& =  \ \ {\rm finite} &  \label{cancel}
\end{eqnarray}
(The Coulomb term is
dominated by the region of asymptotically small momentum transfer
$y \sim x$). This depends only upon the
endpoint behaviour of $f_2$ and leads to (\ref{trans}), as was first shown by
t' Hooft in large-$N_c$ two-dimensional QCD \cite{hoof} (for which only
$f_2$ exists). The same 
endpoint behaviour can be deduced under similar assumptions for any
$f_n$ when one of its arguments vanishes. The usual 
justification given for ignoring pair production (see e.g. \cite{horn})
is based on the assumption
that $f_n$ does not diverge if three or more momenta vanish. In this case,
it is easy to see from (\ref{pair1})(\ref{pair2}) by power counting
that pair production would be subleading as $x \to 0$. 

In fact, the ansatz (\ref{ans}) with (\ref{trans}) fails in general 
to account correctly for the behaviour
of $f_n$ when one momentum vanishes, except for $f_2$ in the valence
approximation. Furthermore,
(\ref{ans}) fails to account for the correct behaviour as two or more
momenta vanish in $f_n$.
This follows by examining the behaviour of eqs.(\ref{mass})-(\ref{four})
when two or more momenta vanish.
To illustrate these points, let us perform a one-loop 
calculation in $e/m$.
In this case, it is easy to see that to leading order in $e/m$,
the solution to the 
four-parton equation (\ref{four}) is
\be
f_4 (x_1, x_2, x_3, x_4) = {e^2 \over 2 \pi} { A(x_1, x_2, x_3)
  \over M^2 - m^2 \sum_{i=1}^{4} {1 \over x_i}}  \label{sol}
\eq
(Note that $M^2 \sim O(m^2)$.) This shows that as two momenta vanish, say
$x_1$  and $x_3$, such that $\lim_{z \to 0} x_1,x_3 \propto z$,
then $f_4$ is non-zero and finite, which is
incompatible with (\ref{ans}). Furthermore, as three momenta vanish,
say $\lim_{z \to 0} x_1, x_2, x_3 \propto z$, 
then $f_4$ diverges as $z^{\b -1 }$ 
if $f_2(z,1-z) \sim z^{\b}$, which is also incompatible with
(\ref{ans}). Therefore, rather than vanishing at the
corners of phase space, we find relations between 
wavefunction components in each Fock sector, showing in particular
that the valence
and the sea are not analytically independent.
Substituting $f_4$ (\ref{sol}) into the two-parton 
equation (\ref{mass})-(\ref{pair2}), one obtains an effective 
valence equation. In the limit of small $x$ the singular part of this
equation reads
\begin{eqnarray}
&\lim_{x \to 0} \left\{ 
{e^2 \over \pi} \int_{0}^{1} dy {f_2(x,1-x) - f_2(y,1-y) \over (x-y)^2}
\left(1 - {e^2 \over 2  m^2 \pi} I(x/y) \right)  \right. & \nonumber \\
 \ \ \  & \left. \! \! \! \! \! \! \! \! \! \! \! \! \! \! 
 + \ {m^2 \over x}f_2(x,1- x) \right\}
& \! \! \! \! \! \! \! \! \! \! \! \! \! \! \! \! \! \! \! \! \! 
\! \! \! \! \! \! \! \! \! \! \! \! \! \! \! \! \! \! \! \! \! 
\! \! \! \! \! \! \! = \ {\rm finite} 
\end{eqnarray}
\begin{eqnarray}
I(w) & = & \int_{0}^{1} dv \ \left\{ {1 \over {1 \over v(1-v)} +w -1}
\left( 1 - {(1-w)^2  \over (1-v(1-w))^2} \right) \right. \nonumber \\
&& \ \ \ \ \ \ \ \ \left. + {1 \over \left( 1 + {w \over v(1-w)} \right)^2
\left( w + {1 \over 1-v} \right)} \right\}
\end{eqnarray}
Comparing with (\ref{cancel}), this 
demonstrates that the neglect of pair production is unjustified in the
determination of $\b$, and that eq.(\ref{trans}) will not be correct
in general.

To summarize, the ansatz (\ref{ans}) with (\ref{trans}) 
correctly accounts for the behaviour of $f_n(x_1 , \dots x_n)$ as
one $x_i \to 0$ only in the  approximation which sets
$f_{m} = 0$ for $m > n$. More generally $\b$ will be renormalised
from the value given by (\ref{trans}). Eq.(\ref{ans}) fails to 
account for the behaviour as two or more momenta vanish.

First order $e/m$ perturbation theory and a 
$4$-particle Tamm-Dancoff Fock-space truncation
were used above, but it is easy to check that 
the phenomenon is not special to these choices. It follows by power
counting in the boundstate integral equations 
when various momenta $x_i$ are made to vanish, i.e. one does not need to 
solve the equations to reach the conclusions. 
Moreover, it is also easy to check that since it only depends
upon the  current-current structure of interactions given in eq.(\ref{ham}),
it will occur in {\em any two-dimensional gauge theory} with pair
production; in particular, for fundamental representations at finite $N_c$
\cite{stan4} and adjoint representations \cite{kleb}. 
In the massive Schwinger model, typically a one percent error in $\b$ can 
lead to a similar size error in $M^2$ estimated from variational ansatz, 
which is significant at the level of precision of current computations
\cite{perry1,koji}. The error is small in this case because the valence
approximation becomes exact in the large and small $m$ limits. In cases
where the latter is not true, such as  boson or Majorana fermion
representations, the effects can be somewhat larger, especially for small
$m$.

The algebraic manipulations have a simple underlying physical explanation.
Higher Fock states renormalise the Coulomb coupling at asymptotically
small momentum transfer through
vacuum polarization and vertex renormalisation. In general,
the parton mass $m^2$ may also be renormalised. These (sector-dependent)
renormalisations
are well-known in the Tamm-Dancoff treatment of light-cone boundstate
integral equations \cite{perry3}. However, it does not seem to have
been generally recognised that formulae such as
(\ref{trans}) are subject to renormalisation as a result.
Although a suitable variational ansatz may 
just include (renormalised) $\b$ as
one of the variational parameters, an ansatz which correctly
incorporates the behaviour as two or more momenta vanish is not so
simple.

A significant effect of these renormalisations might be expected in 
theories which exhibit complete screening in some limit. This is the case
for $m \to 0$ in two-dimensional QED, for example. 
In this limit the string tension is
known to  vanish as $\sigma \sim e m$ \cite{cole}.\footnote{This result 
was established by bosonization.
In principle there can be a renormalisation
of the parameter $m/e$ in the bosonization approach relative
to a fermionic basis, which prevents direct comparison
with light-cone results in such a basis. To the author's knowledge, the
result has not been established in light-cone quantisation in the
fermionic basis independently
of  the groundstate mass $M$ obtained from variational argument using
(\ref{trans}).}
Pair production effects now completely 
screen the linear potential as $m \to 0$, resulting in the Higgs
mechanism and the production of Schwinger's massive photon in the
physical spectrum. 
If we assume a similar renormalisation, in the boundstate integral equations,
of the Coulomb coupling at 
asymptotically small 
momentum transfer $e^2 \to e^{2}_{R} \sim e m$ (this is not entirely
justified if the sources are not heavy), one naively finds
from (\ref{trans}) that $\b \sim \sqrt{m}$ as $m \to 0$, 
rather than $\b \sim m$. 
In any case, 
since it is $e_R$ which is relevant to the determination of $\b$, rather
than $e$, this emphasizes the need to properly account for 
renormalisation effects due
to higher Fock states, before attempting to accurately solve light-cone
gauge theories by variational ansatz. It may have some bearing on the
small discrepancies in results 
found between bosonization and light-cone quantisation of two-dimensional
QED \cite{koji}.

\section{Four-Dimensional Gauge Theories.}

The considerations above are relevant in QCD to the behaviour of partons 
carrying small light-cone momentum fraction $x$ of the hadron. 
Let us first consider non-perturbative
formulations. The transverse lattice colour-dielectric formulation 
of pure QCD is 
conceptually similar to two-dimensional gauge theories with adjoint
boson representations \cite{bard}. 
This is especially true in the large-$N_c$ limit, when an {\em exact}
 dimensional
reduction to two-dimensional gauge theory takes place \cite{us}. 
Use of variational wavefunctions in some shape or form seems
unavoidable to obtain satisfactory numerical convergence with this method.
The job of implementing the subtleties described in the previous section
for higher Fock states looks quite formidable. However,
a hybrid method introduced in ref.\cite{us}, which combines a variational
approach with discretized light-cone quantisation (DLCQ) \cite{cris}, 
offers a compromise.
In this method, the parton momentum fractions $x_i= t_i /K$ are artificially
discretized into integers $t_i$, such that
$\sum_{i=1}^{n} t_i = K$.
When computing matrix elements of $P^-$ in Fock space,
spectator partons retain their discrete momentum $t_i$, while partons
involved in an elementary 
interaction have their momenta  projected onto a continous-momentum
wavefunction basis (see Appendix C of ref.\cite{us} for details), for
evaluation by variational ansatz. For a
theory with elementary 
interactions of order $\leq q$ in the number of fields, no 
more than $q$ partons are involved in such interactions in initial plus
final states. This means that
that an appropriate variational basis need only be found for a few
partons, however many higher Fock states are included in the calculations.
We are currently incorporating the correct
behaviour of variational wavefunctions, discussed in the last section,
into our calculations on the transverse lattice \cite{us2}.
Other variational 
approaches to light-cone QCD break gauge invariance explicitly by
giving large constituent masses to quarks and gluons \cite{perry2}. It is
hoped that higher Fock states are strongly suppressed as a result and that
perturbative renormalisation theory can be performed. 

Finally, it should be emphasized 
that the singular behaviour of light-cone wavefunctions, 
when two or more light-cone momentum fractions $x_i$ vanish, 
has observable implications for the small Bjorken-$x$ region
of hadronic structure functions. This behaviour is determined from the
high-energy boundary condition on light-cone wavefunctions in QCD
\cite{stan2},
which can be easily solved perturbatively and
again relates Fock space sectors with differing numbers of partons. The 
definition of the distribution function for a parton of momentum
fraction $x_1$ is
\begin{eqnarray}
Q(x_1) & = & \sum_{n} 
\int [d{\bf k}] \int dx_2\cdots dx_{n} \delta \left (1-\sum_{i=1}^{n}
x_i \right)  \ \ \ \ \ \ \ \  \nonumber \\
&& \ \ \ \ \ \ \times |f_n(x_1, x_2 , \ldots, x_n)|^2 \ .
\end{eqnarray}
$[d{\bf k}]$ indicates the integration over all parton transverse momenta
upon which $f_n$ also depends (this is absent in two dimensions).
Generalising the analysis of the previous section to all $n$, one finds
that the wavefunctions of two-dimensional
gauge theories diverge slower than $x^{-(s-2)}$  when $s>2$ light-cone
momenta are of order $x \to 0$, and are otherwise finite.
By power counting, this means that they are not singular enough 
to prevent $Q(x \to 0)$ from vanishing in two dimensions.
In four-dimensional gauge theories however, it is possible (via the
quark--gluon--anti-quark vertex for example) for $f_n$ 
to diverge already  when only two parton momenta are vanishing, 
leading to  $\log{x}$
contributions to $Q(x \to 0)$ \cite{stan2,muller}. The accumulation 
of such logarithms
is thought to lead to the power law rise (Regge behaviour) 
exhibited by hadronic structure functions at small $x$.

\vspace{10mm}

\noindent
I would like to thank B. van de Sande for discussions. This work is supported
by a CERN fellowship.

\end{document}